\documentstyle[aps,psfig,multicol]{revtex}
\begin{document}
\setlength{\topmargin}{-1.3cm}
\draft
\title{Correlations in Scale-Free Networks: Tomography and Percolation}
\author{R. Xulvi-Brunet$^1$, W. Pietsch$^1$, and I.M. Sokolov$^{1,2}$}
\address{$^1$Institut f\"{u}r Physik, Humboldt Universit\"{a}t zu Berlin, 
         Invalidenstra\ss e 110, D-10115 Berlin, Germany}
\address{$^2$Theoretische Polymerphysik, Universit\"at Freiburg, 
         Hermann Herder Str. 3, D-79104 Freiburg, Germany}
\maketitle

\bigskip

\begin{abstract}
We discuss three related models of scale-free networks with the same degree
distribution but different correlation properties. Starting from the
Barabasi-Albert construction based on growth and preferential attachment we
discuss two other networks emerging when randomizing it with respect to
links or nodes. We point out that the Barabasi-Albert model displays
dissortative behavior with respect to the nodes' degrees, while the
node-randomized network shows assortative mixing. These kinds of
correlations are visualized by discussig the shell structure of the networks
around their arbitrary node. In spite of different correlation behavior, all
three constructions exhibit similar percolation properties.
\end{abstract}


\begin{multicols}{2}

\section*{Introduction}

Scale-free networks, i.e. networks with power-law degree          
distributions, have recently been widely studied (see Refs.
\cite{baretalb,dorogov} for a review). Such degree distributions have 
been found in many different contexts, for example in several technological 
webs like the Internet \cite{int,past}, the WWW \cite{www2,WWW}, or 
electrical power grids \cite{Wa}, in natural networks like the network of 
chemical reactions in the living cell \cite{Oltvai,Fell,Mason}  
and also in social networks, like the network of human sexual contacts 
\cite{sex}, the science \cite{New1,New2} and the movie actor \cite{Am,Alb} 
collaboration networks, or the network of the phone calls \cite{phonecall}. 

The topology of networks is essential for the spread of information or 
infections, as well as for the robustness of networks against intentional 
attack or random breakdown of elements.
Recent studies have focused on a more detailed topological characterization 
of networks, in particular, in the degree correlations among nodes 
\cite{past,Newm,Berg,Egui,Bo,mas,Vaz,Goh,Bog,NE,Serr}. For instance, many 
technological and biological networks show that nodes with high degree 
connect preferably to nodes with low degree \cite{past,mas}, a property 
referred to as disassortative mixing. On the other hand, social networks 
show assortative mixing \cite{Newm,NE}, i.e. highly connected nodes are 
preferably connected to nodes with high degree.

In this paper we shall study some aspects of this topology, specifically
the importance of the degree correlations, in three related 
models of scale-free networks and concentrate on the two important 
characteristics: the tomography of shell structure around an arbitrary 
node, and percolation.

\section*{The Models}

Our starting model is the one of Barabasi and Albert (BA) \cite{BA-model},
based on the growth algorithm with preferential attachment. Starting from an
arbitrary set of initial nodes, at each time step a new node is added to the
network. This node brings with it $m$ proper links which are connected to $m$
nodes already present. The latter are chosen according to the preferential
attachment prescription: The probability that a new link connects to a
certain node is proportional to the degree (number of links) of that node.
The resulting degree distribution of such networks tends to 
\cite{Redner,degdis,Kra}: 
\begin{equation}
P(k)=\frac{2m(m+1)}{k(k+1)(k+2)} \sim k^{-3}.  \label{degdis}
\end{equation}
Krapivsky and Redner \cite{Kra} have shown that in the BA-construction 
correlations develop spontaneously between the degrees of connected nodes.
To assess the role of such correlations we shall randomize the BA-network.

Recently Maslov and Sneppen \cite{mas} have suggested an algorithm radomitzing
a given  network that keeps the degree distribution constant. According 
to this algorithm at each step two links of the network are chosen at random.
Then, one end of each link is selected randomly and the attaching nodes 
are interchanged. However, in case one or both of these new links already 
exits in the network, this step is discarded and a new pair of edges is 
selected. This restriction prevents the apparearance of multiple edges 
connecting the same pair of nodes. A repeated application of the 
rewiring step leads to a randomized version of the original 
network. We shall refer to this model as link-randomized (LR) model. 

The LR model can be compared with another model which is widely studied 
in the context of scale-free networks, namely with 
the configuration model introduced by Bender and Canfield 
\cite{cand,mollreed}. It
starts with a given number $N$ of nodes and assigning to each node a number 
$k_i$ of ``edge stubs'' equal to its desired connectivity. The stubs of 
different nodes are then connected randomly to each other; two connected 
stubs form a link. One of the limitations of this ``stub reconnection'' 
algorithm is that for broad distribution of connectivities, which is 
usually the case in complex networks, the algorithm generates multiple
edges joining the same pair of hub nodes and loops connecting the node to 
itself. However, the cofiguration model and the LR model get equivalent as 
$ N\rightarrow \infty $.
 
One can also consider a node-randomized (NR) counterpart of the LR 
randomize procedure. The only difference to the link-radomized algorithm
is that instead of choosing randomly two links we choose randomly two nodes 
in the network. Then the procedure is the same as in the LR model.

As we proceed to show, the three models have different properties with
respect to the correlations between the degrees of connected nodes. While
the LR (configuration) model is random, the genuine BA prescription leads to
a network which is dissortative with respect to the degrees of connected
nodes, and the NR model leads to an assortative network. This fact leads to
considerable differences in the shell structure of the networks and also to
some (not extremely large) differences in their percolation characteristics.
We hasten to note that our simple models neglect many important aspects of real
networks like geography \cite{Soki,geog} but stress the importance to
consider the higher correlations in the degrees of connected nodes.

\section*{Tomography of the Networks}

\label{shell structure}Referring to spreading of computer viruses or human
diseases, it is necessary to know how many sites get infected on each step
of the infection propagation. Thus, we examine the local structure in the
network. Cohen et al. \cite{tomography} examined the shells around the node
with the highest degree in the network. In our study we start from a node
chosen at random. This initial node (the root) is assigned to shell number
0. Then all links starting at this node are followed. All nodes reached are
assigned to shell number 1. Then all links leaving a node in shell 1 are
followed and all nodes reached that don't belong to previous shells are
labelled as nodes of shell 2. The same is carried out for shell 2 etc.,
until the whole network is exhausted. We then get $N_{l,r}$, the number of nodes
in shell $l$ for root $r$. The whole procedure is repeated starting at all $%
N $ nodes in the network, giving $P_{l}(k)$, the degree distribution in
shell $l$. We define $P_{l}(k)$ as:
\begin{equation}
   P_{l}(k)=\frac{\sum_{r} N_{l,r}(k)}{\sum_{k,r} N_{l,r}(k)}. \label{aa}
\end{equation}

 We are most interested in the average degree $\langle k\rangle
_{l}=\sum_{k}kP_{l}(k)$ of nodes of the shell $l$. In the
epidemiological context, this quantity can be interpreted as a disease
multiplication factor after $l$ steps of propagation. It describes how many
neighbors a node can infect on average. Note that such a definition of 
$P_{l}(k)$ gives us for the degree distribution in the first shell:
\begin{equation}
   P_{1}(k)=\frac{\sum_{r} N_{1,r}(k)}{\sum_{k,r} N_{1,r}(k)}=
   \frac{kN_k}{\sum_k kN_k}=\frac{kP(k)}{\langle k \rangle},  \label{bb}
\end{equation}
where $P(k)$ and $N_k$ are the degree distribution and the number of nodes 
with degree $k$ in the network respectively. We bear in mind that every link 
in the network is followed exactly once in each direction. Hence, we find 
that every node with degree $k$ is counted exactly $k$ times. From 
Eq.($\ref{bb}$) follows that 
$\langle k\rangle _{1}=\langle k^{2}\rangle / \langle k\rangle$.
This quatity, that plays a very important role in the percolation theory of
networks \cite{cohetal}, depends only on the first and second moment of the
degree distribution, but not on the correlations. Of course $P_0(k)=P(k)$.

Note that as $N\rightarrow \infty $ we have $\langle k^{2}\rangle
\rightarrow \infty $: for our scale-free constructions the mean degree in
shell 1 depends significantly on the network size determining the cutoff in
the degree distribution. However, the values of $\langle k\rangle _{1}$ are
the same for all three models: The first two shells are determined only by
the degree distributions. In all other shells the three models differ. For
the LR (configuration) model one finds for all shells in the thermodynamic 
limit $P_{l}(k)=P_{1}(k)$. However, since these distributions do not
possess finite means, the values of $\langle k\rangle _{l}$ are governed
by the finite-size cutoff, which is different in different shells,
since the network is practically exhausted within the first few steps, see 
Fig.1.

\begin{figure}[tbp]
  \centerline{\hspace{0cm}\psfig{figure=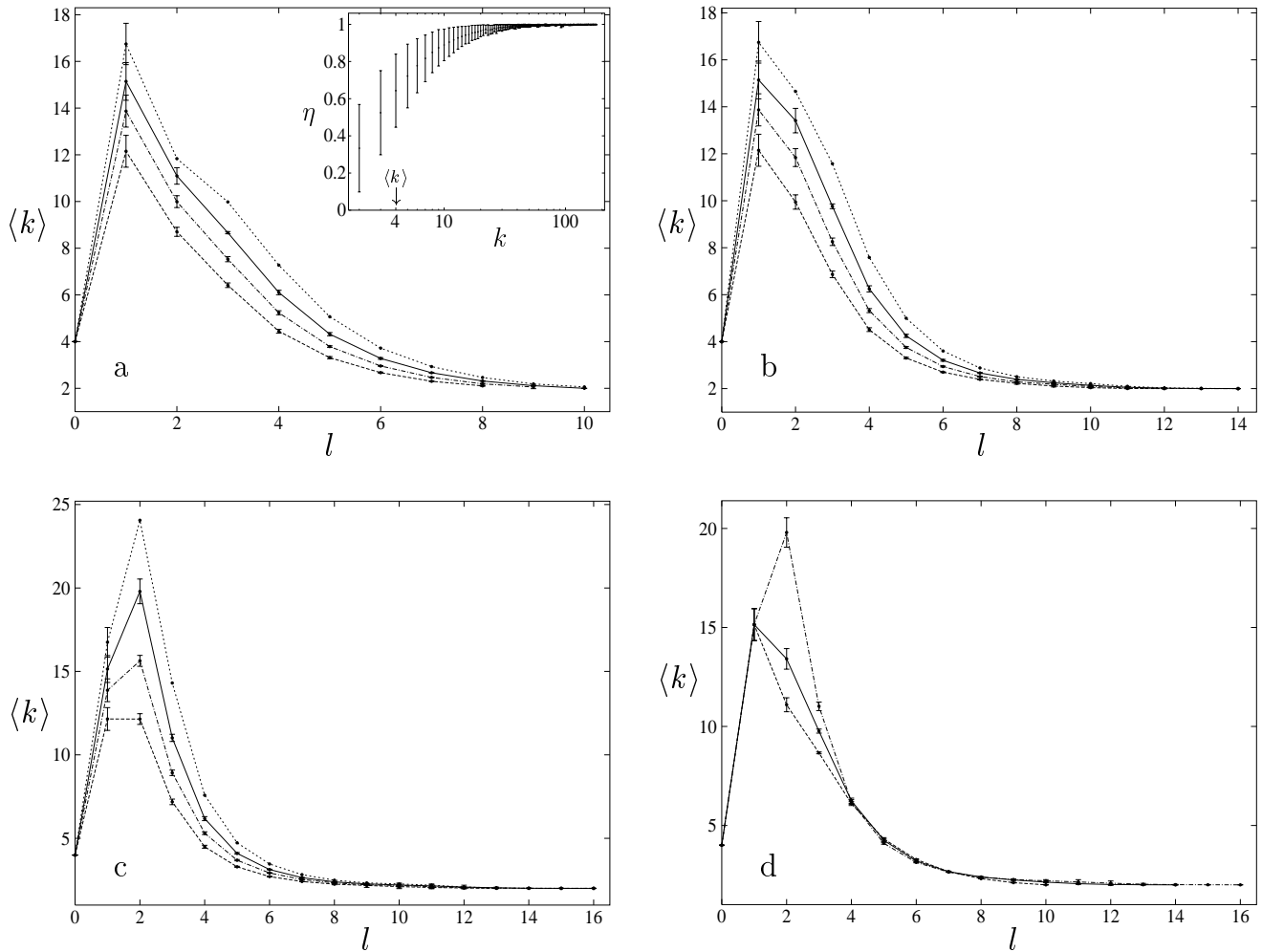,width=7.1in}}
    \caption{Mean degree value $\langle k\rangle $ in shell $l$: (a) for 
             the BA-model, 
             (b) for the LR-model, (c) for the NR-model. Different curves 
             correspond to different network sizes: from top to bottom 
             100,000; 30,000; 10,000; 3,000 nodes. 10 simulations were done
             for each value except for the shells with $l\geq 2$ at 
             $N=100,000$ based on only one simulation. Panel (d) compares 
             the tomography of the models with $N=30,000$: from top to 
             bottom NR model; LR model; BA model. The inset in panel (a) shows
            the average age $\eta $ of a node as a function of its degree $k$.}
    \label{fig1}
\end{figure}

In what follows we compare the shell structure of the BA, the LR and the NR
models. We discuss in detail the networks based on the BA-construction with 
$m=2$. For larger $m$ the same qualitative results were observed. In the 
present work we refrain from discussion of a peculiar case $m=1$. For $m=1$ 
the topology of the BA-model is distinct from one for $m\geq 2$ since in this 
case the network is a tree. This connected tree is destroyed by the 
randomization procedure and is transformed into a set of disconnected 
clusters. On the other hand, for $m\geq 2$ the creation of large separate 
clusters under randomization is rather unprobable, so that most of the nodes 
stay connected. Fig. \ref{fig1} shows $\langle k\rangle $ as a function of 
the shell number $l$. Panel (a) corresponds to the BA model, panel (b) to 
the LR model, and panel (c) to the NR model. The different curves show 
simulations for different network sizes: $N=3,000$; $N=10,000$; $N=30,000$; 
and $N=100,000$. All points are averaged over ten different realizations 
except for those for networks of 100,000 nodes with only one simulation. In 
panel (d) we compare the shell-structure for all three models at $N=30,000$. 
The most significant feature of the graphs is the difference in 
$\langle k \rangle _{2}$. In the BA and LR models the maximum is reached in 
the first shell, while for the NR model the maximum is reached only in the 
second shell: 
$\langle k\rangle_{2,BA}<\langle k\rangle _{2,LR}<\langle k\rangle _{2,NR}$. 
This effect becomes more pronounced with increasing network size. In shells 
with large $l$ for all networks mostly nodes with the lowest degree $2$ are 
found.

The inset in graph (a) of Fig. \ref{fig1} shows the relation between 
average age $\eta$ of nodes with connectivy $k$ in the network as a function 
of their degree for the BA model. The age of a node $n$ and of any of its 
proper links is defined as $\eta (n)=(N-t_{n})/N$ where $t_{n}$ denotes the 
time of birth of the node. For the randomized LR and NR models age has no 
meaning. The figure shows a strong correlation between age and degree of a 
node. The reasons for these strong correlations are as follows:
First, older nodes experienced more time-steps than younger ones and thus
have larger probability to acquire non-proper bonds. Moreover, at earlier
times there are less nodes in the network, so that the probability of
acquiring a new link per time step for an individual node is even higher.
Third, at later time-steps older nodes already tend to have higher degrees
than younger ones, so the probability for them to acquire new links is
considerably larger due to preferential attachment. The correlations between
the age and the degree bring some nontrivial aspects into the BA model based 
on growth, which are erased when randomizing the network.

Let us discuss the degree distribution in the second shell. In this case we
find as that every link leaving a node of degree $k$ is counted $k-1$ times. 
Let $P(l|k)$ be a probability that a link leaving a node of degree $k$ 
enters a node with degree $l$. Neglecting the possibility of short loops 
(which is always appropriate in the thermodynamical limit 
$N \rightarrow \infty$) and the inherent direction of links (which may be 
not totally appropriate for the BA-model) we have: 
\begin{equation}
P_{2}(l)=\frac{\sum_k kP(k)(k-1)P(l|k)}{\sum_{k}kP(k)(k-1)}.  \label{P2}
\end{equation}

\end{multicols}

\bigskip

\begin{multicols}{2}

The value of $\langle k\rangle _{2}$ gives important information about the
type of mixing in the network. To study mixing in networks one needs to
divide the nodes into groups with identical properties. The only relevant
characteristics of the nodes that is present in all three models, is their
degree. Thus, we can examine the degree-correlations between neighboring
nodes, which we compare with the uncorrelated LR model, where the
probability that a link connects to a node with a certain degree is
independent from whatever is attached to the other end of the link: 
$P(k | l)=kP(k)/\langle k\rangle =kP(k)/2m$. All other relations would 
correspond to assortative or disassortative mixing. Qualitatively, 
assortativity then means that nodes attach to nodes with similar degree more 
likely than in the LR-model: 
$P(k | l)>P(k | l)_{LR}=kP(k) / \langle k\rangle$ for $k\approx l$. 
Dissortativity means that nodes attach to nodes with very different degree 
more likely than in the LR-model: $P(k | l)> kP(k) / \langle k\rangle $ 
for $k\gg l$ or $l\gg k$. Inserting this in Eq.(\ref{P2}), and calculating 
the mean, one finds qualitatively that 
$\langle k \rangle _{1}=\langle k\rangle _{2,LR}<\langle k\rangle _{2}$ for
assortativity, and $\langle k\rangle _{1}>\langle k\rangle _{2}$ for
dissortativity.

In the following we show where the correlations of the BA and NR model
originate. A consequence of the BA-algorithm is that there are two different
types of ends for the links. Each node has exactly $m$ proper links
attached to it at the moment of its birth and a certain number of links that
are attached later. Since each node receives the same number of links at
its birth, towards the proper nodes a link encounters a node with degree $k$
with probability $P(k)$. To compensate for this, in the other direction a
node with degree $k$ is encountered with the probability $\frac{(k-m)P(k)}{m}
=2\frac{kP(k)}{\langle k\rangle }-P(k)$, so that both distributions together
yield $kP(k)/\langle k\rangle $. On one end of the link nodes with small
degree are predominant: $P(k)<kP(k)/\langle k\rangle $ for small $k$. On the
other end nodes with high degree are predominant: $(k-m)P(k)/m>kP(k)/2m$ for 
$k$ large. This corresponds to dissortativity. Actually the situation is
somewhat more complex since in the BA model these probability distributions
also depend on the age of the link.

Assortativity of the NR model is a result of the node-randomizing process.
Since the nodes with smaller degree are predominant in the node population,
those links are preferably chosen that have on the end with the randomly
chosen node a node with a smaller degree ($P(k)>kP(k)/\langle k\rangle $ for 
$k$ small). Then the randomization algorithm exchanges the links and
connects those nodes to each other. This leads to assortativity for nodes
with small degree, which is compensated by assortativity for nodes with high
degree.

\section*{Percolation}

Percolation properties of networks are relevant when discussing their
vulnerability to attack or immunization which removes nodes or links from
the network. For scale-free networks random percolation as well as
vulnerability to a deliberate attack have been studied by several groups 
\cite{cohetal,ben,je,cohetal2,callnew}. One considers the removal of a
certain fraction of edges or nodes in a network. Our simulations
correspond to the node removal model; $q$ is the fraction of removed nodes. 
Below the percolation threshold $q<q_{c}$ a giant component (infinite 
cluster) exists, which ceases to exist above the threshold. A giant 
component, and consequently $q_{c}$ is exactly defined only in the 
thermodynamic limit $ N\rightarrow \infty $: it is a cluster, to which 
a nonzero fraction of all nodes belongs.

In \cite{mollreed} and \cite{cohetal} a condition for the percolation
transition in random networks has been discussed: Every node already
connected to the spanning cluster is connected to at least one new node.
Ref. \cite{cohetal} gives the following percolation criterion for the
configuration model: 
\begin{equation}
1-q_{c}=\frac{\langle k\rangle }{\langle k^{2}\rangle -\langle k\rangle },
\label{condperc}
\end{equation}
where the means correspond to an unperturbed network ($q=0$). For networks
with degree distribution Eq.($\ref{degdis}$), $\langle k^{2}\rangle $ 
diverges as $N \rightarrow \infty$. This yields for the random networks 
with a such degree distribution a percolation threshold $q_{c}=1$ in the 
thermodinamic limit, independent of the minimal degree $m$; in the 
epidemiological terms this corresponds to the absence of herd 
immunities in such systems. Crucial for this threshold is the power-law 
tail of the degree distribution with an exponent $\leq 3$. Moreover, Ref. 
\cite{ben} shows that the critical exponent $\beta $ governing the 
fraction of nodes $M_{\infty }$ of the giant component, 
$M_{\infty }\propto (q_{c}-q)^{\beta }$, diverges as the exponent of the 
degree distribution approaches $-3$. Therefore $M_{\infty }$ approaches zero 
with zero slope as $q\rightarrow 1$.

\begin{figure}[tbp]
   \centerline{\hspace{0cm}\psfig{figure=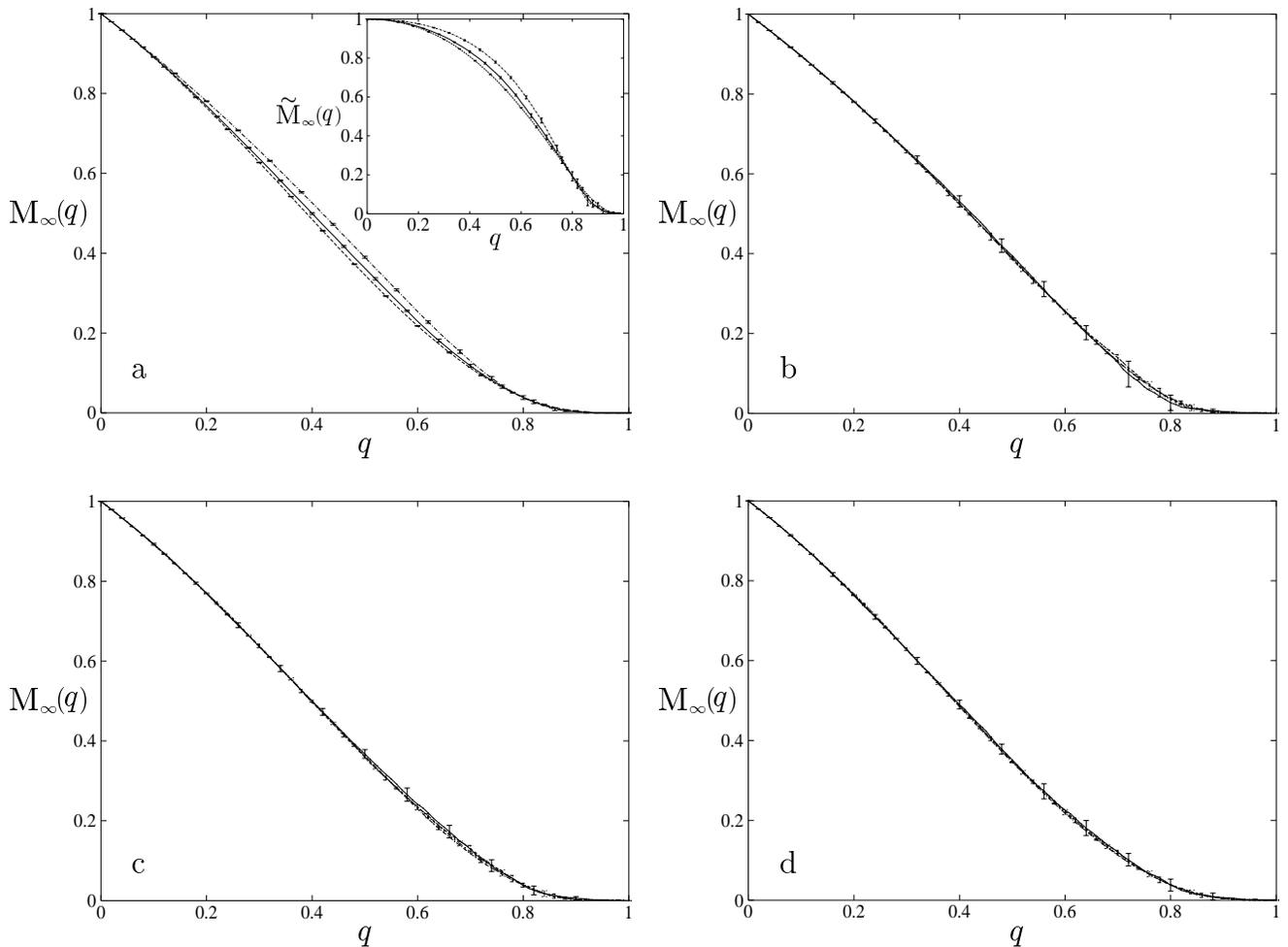,width=7.1in}}
      \caption{Fraction of nodes $M_{\infty }$ in the giant component 
               depending on the fraction $q$ of nodes removed from the 
               network: (b) for the BA-model, (c) for the LR-model, (d) 
               for the NR-model. Different curves correspond to different 
               network sizes: from top to bottom 300,000 (5 simulations); 
               30,000; 10,000; 3,000 nodes (10 simulations each). Graph (a) 
               compares all three models at $N=300,000$ (from top to bottom: 
               BA-model, LR-model, NR-model). The inset shows the fraction 
               $\tilde{M}_{\infty }$ of the number of nodes in the giant 
               component relative to the remaining number of nodes in the
               network $(1-q)N$.}
      \label{fig2}
\end{figure}

In Fig. \ref{fig2} we plotted for the three models discussed $M_{\infty }$ 
as a function of $q$.
The behavior of all three models for a network size of $300,000$ nodes is
presented in panel (a). In the inset the size of the giant component was 
measured in relation to the number of nodes remaining in the network 
$(1-q)N$ and not to their initial number $N$. Other panels show the 
percolation behavior of each
of the models at different network sizes: Panel (b) corresponds to the BA
model, (c) to the LR model, and (d) to the NR model. For the largest
networks with $N=300,000$ nodes we calculated 5 realizations for each model,
for those with $30,000$; $10,000$; and $3,000$ nodes averaging over 10
realization was performed.
For all three models within the error bars the curves at different network
sizes coincide. This shows that even the smallest network is already close
to the thermodynamical limit. R. Albert et al. have found a similar behavior
in a study of BA-networks \cite{je}. They analyze networks of sizes 
$N=1000, 5000$ and $20000$ concluding ``that the overall clustering scenario
and the value of the critical point is independent of the size of the system''.

In the simulations we find two regimes: for moderate $q$ we find, that the 
sizes of the giant components of the BA, LR, and NR model obey the 
inequalities $M_{\infty ,BA}>M_{\infty ,LR}>M_{\infty ,NR}$
, while for $q$ close to unity the inequalities are reverted: $M_{\infty
,BA}<M_{\infty ,LR}<M_{\infty ,NR}$. However, in this regime the differences
between $M_{\infty ,BA},M_{\infty ,LR}$ and $M_{\infty ,NR}$ are subtle and
hardly resolved on the scales of Fig. 2. We note that similar situation
was observed in Ref. \cite{Newm}. However, there the size of 
the giant cluster was measured not as a function of $q$ but of a scaling 
parameter in the degree distribution. 

The observed effects can be explained by the correlations in the network.
For $q=0$ one has $M_{\infty,BA}=M_{\infty ,LR}=M_{\infty ,NR}$. Now,
the probability that single nodes loose their connection to the giant cluster 
depends only on the degree distribution, and not on correlations. So, the 
difference in the $M_{\infty}$ must be explained by the break-off of 
clusters containing more than one node. The probability for such an event 
is smaller in the BA than in the LR model, since dissortativity implies 
that one finds fewer 'regions', where only nodes with low degree are present.

However, when we get to the region of large $q$, as nodes with low degree
act as 'bridges' between the nodes with high degree, the connections
between the nodes with high degree are weaker in the case of the BA model
than in the case of the LR model. So, the probability that nodes with high
degree break off is higher for the BA model than for the LR model. There is
no robust core of high-degree nodes in the network \cite{Newm}. 
The correlation effects for the NR model, when compared with the LR model, 
are opposite to those for the BA model.

\end{multicols}

\bigskip

\begin{multicols}{2}

\section*{Conclusion}

We consider three different models of scale-free networks: the genuine
Barabasi-Albert construction based on growth and preferential attachment,
and two networks emerging when randomizing it with respect to links or
nodes. We point out that the BA model shows dissortative behavior with
respect to the nodes' degrees, while the node-randomized network shows
assortative mixing. However, these strong differences in the shell structure
lead only to moderate quantitative difference in the percolation behavior of
the networks.

\section*{Acknowledgment}

Partial financial support of the Fonds der Chemischen Industrie is gratefully
acknowledged.

\end{multicols}

\end{document}